\documentclass[sigconf,nonacm]{acmart}

\setcopyright{none}
\acmDOI{}
\acmISBN{}
\acmYear{2026}

\acmConference[EARL @ CIKM '26]
  {3rd Workshop on Evaluating and Applying Recommender Systems with Large Language Models (EARL) at CIKM 2026}
  {November 8, 2026}
  {Rome, Italy}

\AtBeginDocument{%
  }

\usepackage{tikz}
\usepackage{array}
\usetikzlibrary{arrows.meta}
\usepackage{microtype}
\begin{document}

\title[Preprocessing Choices in LLM-Based Recommendation]{Preprocessing Choices in Fine-Tuning Data Matter: Utility and Demographic Sensitivity in LLM-Based Recommendation}

\author{Furkan Yasir Goksu}
\authornote{Also with Artificial Intelligence and Data Science Application and Research Center, Istanbul Technical University.}

\affiliation{%
  \department{Department of Artificial Intelligence and Data Engineering}
  \institution{Istanbul Technical University}
  \city{Istanbul}
  \country{Türkiye}
}
\email{goksu20@itu.edu.tr}

\author{Yaren Yilmaz}
\authornotemark[1]
\affiliation{%
  \department{Department of Computer Engineering}
  \institution{Istanbul Technical University}
  \city{Istanbul}
  \country{Türkiye}
}
\email{yilmazy20@itu.edu.tr}

\author{Sule Gunduz Oguducu}
\authornotemark[1]
\affiliation{%
  \department{Department of Artificial Intelligence and Data Engineering}
  \institution{Istanbul Technical University}
  \city{Istanbul}
  \country{Türkiye}
}
\email{sgunduz@itu.edu.tr}

\renewcommand{\shortauthors}{Goksu et al.}

\begin{abstract}
Large language models (LLMs) have introduced a generative approach to recommendation, producing item lists from textual representations of user histories and item information. However, this generative setting raises challenges concerning output validity, prompt consistency, and sensitivity to demographic descriptors. In this work, we investigate how preprocessing choices in fine-tuning data affect recommendation utility and fairness-related behavior in a generative LLM-based recommender system. 
We first establish a controlled recommendation pipeline that combines supervised fine-tuning, catalog-con\-strained decoding, and prompt augmentation. We then study three preprocessing choices: item-title representation, demographic balancing, and auxiliary item-knowledge enrichment. 
Experiments on MovieLens-1M show that the catalog-con\-strained pipeline eliminates out-of-catalog items and ensures complete ten-item outputs, while prompt augmentation improves prompt consistency and utility. Original MovieLens titles increase HR@10 from 10.67\% to 12.47\% and Coverage@10 from 41.36\% to 47.88\%. Fairness-related effects are attribute-dependent: age balancing and knowledge enrichment reduce selected group-level utility gaps while largely preserving utility, whereas gender balancing reduces utility without improving the evaluated gender-related indicators. Overall, preprocessing choices in fine-tuning data should be evaluated jointly in terms of utility, output behavior, and demographic sensitivity.
\end{abstract}



\maketitle
{\renewcommand\thefootnote{}\footnotetext{\textit{3rd Workshop on Evaluating and Applying Recommender Systems with Large Language Models (EARL) at CIKM 2026, November 8, 2026, Rome, Italy.}}\addtocounter{footnote}{-1}}

\section{Introduction}

Large language models (LLMs) are increasingly being used in recommender systems, enabling recommendations through natural language prompts \cite{liang-etal-2025-taxonomy}, conversational interaction \cite{li-etal-2025-chatcrs}, textual item representations \cite{wang2025beyond}, and domain-specific fine-tuning \cite{hou2026reccocktail}.
Their understanding of natural language allows user histories, preferences, and candidate items to be represented in a single format. However, text-based recommendation introduces risks that are less apparent in conventional recommender systems \cite{munson2025review}, including malformed outputs, repeated or out-of-catalog items \cite{liao2026eliminating}, and prompt sensitivity \cite{li2025ganprompt}. These challenges highlight the importance of understanding not only the quality of LLM-generated recommendations, but also the factors that shape their outputs. 

Unlike conventional recommender systems that typically learn user and item representations from interaction data, LLM-based recommender systems can formulate recommendation tasks through natural language and textual representations of user interactions, item information, and recommendation instructions. 
In this setting, the way user preferences, interaction histories, and item information are expressed in the model input can directly influence the generated recommendations. This makes both the design of training data and the formulation of inference-time prompts important design choices. 
In generative LLM-based recommendation settings, the model generates a sequence of recommended items conditioned on a prompt that represents the user's interaction history and may also include demographic attributes, which are commonly treated as sensitive attributes in fairness research. This raises an important concern about whether LLM-based recommender systems can be sensitive to demographic information embedded in the model input. Although end users may not explicitly state demographic attributes in a recommendation request, such attributes may still be available to a recommender through user profiles or other contextual information. We therefore introduce demographic descriptors explicitly as a controlled intervention, allowing us to examine whether recommendation outputs change when demographic information is varied while the interaction history is held fixed. 

Recent studies show that sensitive attributes can influence LLM recommendation outcomes, suggesting that fairness assessment in this setting requires more than standard accuracy metrics \cite{zhang2023chatgptfair,sah2025faireval}. Evaluation should therefore consider not only whether a model recovers held-out preferences, but also whether its outputs remain stable across demographic conditions and whether utility differs among user groups. At the same time, demographic sensitivity by itself does not necessarily constitute evidence of discrimination. We therefore interpret differences in recommendation outcomes across demographic conditions as fairness-related signals rather than direct evidence of discrimination \cite{wu2023fairness,vassoy2024consumer}. Such comparisons, however, require recommendation outputs that are sufficiently structured and consistent across conditions. LLM recommenders do not always provide this level of output control \cite{munson2025review}. In particular, variations in output length or format can make it difficult to determine whether changes in recommendation lists are caused by preprocessing choices or by the generation process itself. This makes output control an important prerequisite for evaluating how preprocessing choices affect recommendation behavior.

This raises an important question of whether demographic sensitivity can be influenced before recommendation generation and, if so, which aspects of the recommendation pipeline affect it. Recent studies on LLM-based recommender systems have explored several adaptation strategies, including prompt-based recommendation, instruction tuning, hybrid textual and behavioral representations, data augmentation, and the use of external item knowledge \cite{bao2023tallrec,liao2024llara,wang2024large,lin2025can}. Together, these studies suggest that recommendation behavior can be shaped by choices made at different stages of the LLM recommendation pipeline.
In parallel, fairness-oriented work has examined whether LLM recommenders produce different outcomes across sensitive user attributes and how such behavior can be mitigated through prompt- or inference-level interventions \cite{zhang2023chatgptfair,hua2024up5,fayyazi2025facter}. 
%
Despite these advances, comparatively less attention has been paid to the construction of fine-tuning data itself as a design factor and to how such choices may jointly affect recommendation utility and fairness-related behavior. 
In this work, we investigate three preprocessing choices in fine-tuning data, focusing on how item titles are represented, how demographic groups are distributed across the fine-tuning data, and whether additional item information is incorporated through auxiliary fine-tuning examples. We study these factors in the context of catalog-con\-strained LLM recommendation, without assuming that preprocessing alone resolves fairness concerns.

This study examines preprocessing in fine-tuning data as a design factor that can affect recommendation utility, output control, and demographic sensitivity together.  
We address four research questions. We first establish whether the generative recommendation pipeline provides a sufficiently controlled basis for comparison (RQ1), and then examine three preprocessing choices in fine-tuning data concerning item representation, demographic balancing, and auxiliary item knowledge (RQ2–RQ4).
\begin{itemize}
    \item RQ1: Does the recommendation pipeline provide sufficient catalog validity, output control, and prompt consistency for controlled comparisons of preprocessing choices in fine-tuning data?
    \item RQ2: How does item-title representation affect recommendation utility and catalog usage within the catalog-con\-strained recommendation pipeline?
    \item RQ3: How does demographic balancing in fine-tuning data affect recommendation utility and sensitivity to demographic descriptors?
    \item RQ4: Can knowledge-enriched fine-tuning data reduce demographic sensitivity while preserving recommendation utility?
\end{itemize}
We investigate these questions by fine-tuning multiple LLM-based recommender variants within a shared catalog-con\-strained decoding framework. 
The variants differ only in three preprocessing choices in fine-tuning data: item-title representation, demographic balancing of recommendation examples, and auxiliary item-knowledge enrichment. We evaluate all variants under a shared protocol that combines utility with behavioral and fairness-oriented diagnostics. We measure whether generated lists recover held-out user preferences, remain valid and diverse within the target catalog, and change when demographic descriptors are added to the prompt.
Evaluating these dimensions together reveals trade-offs that accuracy alone would miss, particularly when models achieve similar recommendation quality but exhibit different levels of demographic sensitivity.

Our findings demonstrate that preprocessing choices in fine-tuning data have heterogeneous effects on both utility and demographic sensitivity. Item-title representation substantially affects recommendation utility and catalog usage, while demographic balancing produces attribute-dependent outcomes: age balancing improves some fairness-related indicators while largely preserving recommendation utility, whereas gender balancing does not provide the same benefit. Auxiliary item-knowledge enrichment reduces the gender-related SNSR and narrows the neutral group-level utility gap for both attributes, although its effect on the remaining indicators is attribute-dependent. These interventions do not eliminate fairness concerns, but they provide empirical evidence that preprocessing choices in fine-tuning data can influence the trade-off between recommendation utility and fairness-related behavior.

The contributions of this work are as follows:
\begin{itemize}
    \item We establish a controlled catalog-con\-strained LLM recommendation framework with prompt augmentation that provides a consistent experimental backbone for comparing preprocessing choices in fine-tuning data. 
    \item We study three preprocessing choices in fine-tuning data: how items are presented, how demographic groups are distributed in the fine-tuning data, and whether additional item information is introduced through auxiliary fine-tuning examples.  We analyze their effects on recommendation utility, group-level utility differences, and sensitivity to demographic descriptors, without introducing an explicit fairness-specific training objective.


    \item We employ a unified evaluation protocol for catalog-con\-strained LLM recommendation that jointly considers recommendation utility, output-control behavior, group-level utility differences, and demographic prompt sensitivity.
    \item We provide empirical evidence that preprocessing in fine-tuning data is not merely a technical preparation step, but a consequential design choice that can affect both recommendation performance and fairness-related behavior in LLM-based recommender systems.
\end{itemize}

\section{Literature Review}

\textit{Generative Recommendation and Catalog Grounding.} Many LLM-based recommender systems represent users, items, collaborative interaction histories, and recommendation targets as textual sequences. 
Among the foundational approaches, P5 proposed a unified text-to-text framework for multiple recommendation tasks, followed by systems such as OpenP5 for developing and evaluating generative recommenders and GPTRec for autoregressive sequential recommendation, and GenRec, which formulates sequential recommendation as a sequence-to-sequence generation task
\cite{geng2022p5,xu2024openp5,petrov2023gptrec,cao2024genrec}. Subsequent work has further explored instruction tuning, collaborative signals, data augmentation, and external knowledge to adapt LLMs to recommendation \cite{bao2023tallrec,liao2024llara,wang2024large,xu2025rallrec}.

Nevertheless, compared with conventional fixed-candidate ranking, generative recommender systems face challenges such as malformed, repeated, or out-of-catalog items \cite{munson2025review}. Catalog-grounding methods address these issues through retrieval, constrained generation, or item tokenization \cite{liao2026eliminating}. Within such frameworks, item representation is an important design choice, as alternative representation strategies can influence the way items are encoded and generated by the model \cite{xu2024openp5}. Catalog constraints therefore provide a controlled output space for comparing alternative representation choices, although they do not themselves guarantee relevance, diversity, or fairness.

\textit{Prompt Sensitivity and Fairness.}
LLM-based recommendations can be sensitive to variations in prompts and generation, leading to inconsistent outputs for similar recommendation contexts. Recent studies 
show that variations in prompt formulation and repeated generation can affect recommendation outcomes \cite{li2025ganprompt,kweon2025uncertainty}.
This variability is relevant to fairness evaluation, because recommendation changes associated with demographic information should be distinguished from general prompt sensitivity and output differences.

Fairness in recommendation concerns whether recommendation outcomes differ systematically across users or groups, particularly with respect to demographic attributes \cite{wu2023fairness,vassoy2024consumer}. In LLM-based recommendation, recent studies have examined whether sensitive user attributes influence generated recommendations and have proposed fairness-aware prompting and mitigation strategies  \cite{zhang2023chatgptfair,sah2025faireval,hua2024up5,fayyazi2025facter, andre2025bias}. 
Thus, differences in recommendation outcomes across demographic conditions should be interpreted under controlled prompt conditions and together with recommendation utility, rather than as direct evidence of unfairness.

\textit{Preprocessing Choices in Fine-Tuning Data.}
Data-level interventions have been explored in conventional recommender systems as a means of improving fairness. Chen et al. employ data augmentation to rebalance the training distribution, while Melchiorre et al. investigate resampling as a preprocessing strategy for mitigating gender-related disparities in recommendation outcomes \cite{chen2023improving,melchiorre2021investigating}. These studies demonstrate that the composition of training data can itself serve as an intervention point for fairness.

In LLM-based recommendation, prior work has examined aspects of fine-tuning data and task construction primarily from the perspectives of recommendation performance and efficiency. 
Lin et al. investigate representative sample selection for fine-tuning \cite{lin2024data}, while Liu et al. propose a user-level supervised fine-tuning (SFT) formulation for injecting collaborative information into an LLM-based sequential recommendation framework \cite{liu2024p2rec}. However, comparatively less attention has been paid to how preprocessing choices in LLM fine-tuning data affect both recommendation utility and sensitivity to demographic information. 

To address this gap, this study examines three preprocessing choices in the fine-tuning data, namely item title representation, demographic balancing, and auxiliary item-knowledge enrichment, within a shared catalog-con\-strained framework. 

\section{Methodology}

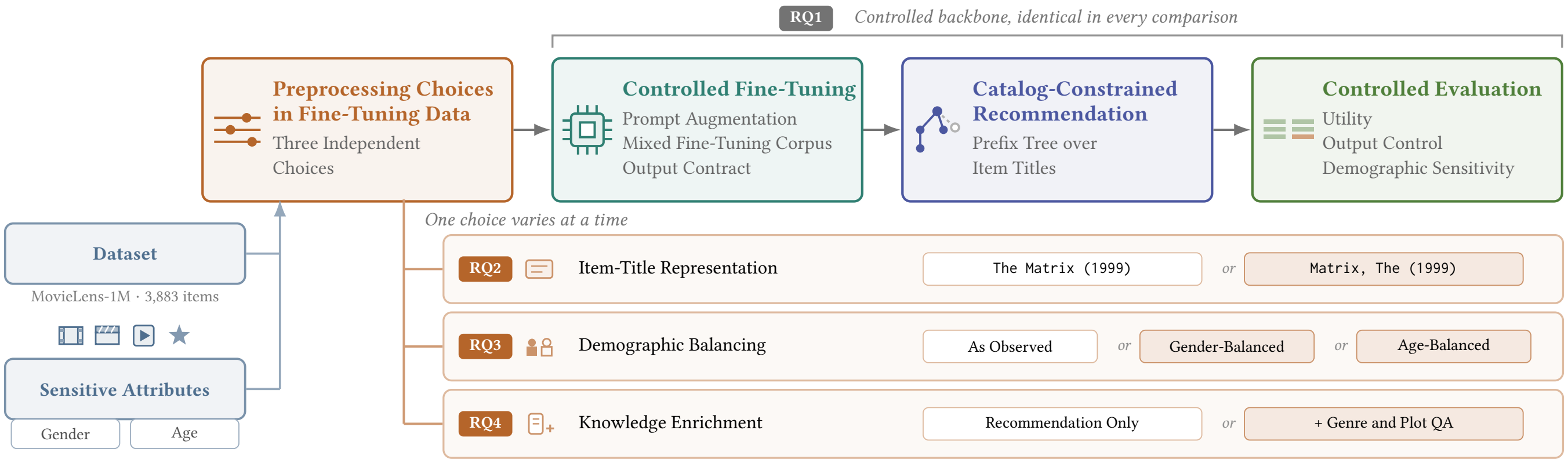
\begin{figure*}[t]
\centering
\definecolor{cData}{HTML}{4E6E8E}
\definecolor{cPrep}{HTML}{B5652A}
\definecolor{cTune}{HTML}{2F7F72}
\definecolor{cDec}{HTML}{4B57A0}
\definecolor{cEval}{HTML}{4F7F3A}
\resizebox{\textwidth}{!}{%
\begin{tikzpicture}[
  stage/.style  = {draw=#1, line width=0.9pt, rounded corners=3pt, fill=#1!8,
                   minimum width=4.00cm, minimum height=1.85cm, anchor=center},
  ftxt/.style   = {anchor=west, inner sep=0pt, font=\footnotesize},
  inbox/.style  = {draw=cData!75, line width=0.8pt, rounded corners=3pt, fill=cData!8,
                   minimum width=3.10cm, minimum height=0.78cm, anchor=center,
                   font=\footnotesize\bfseries, text=cData},
  arr/.style    = {-{Latex[length=2.2mm,width=1.6mm]}, line width=0.85pt, black!55},
  lead/.style   = {line width=0.8pt, cData!60},
  rqrow/.style  = {draw=cPrep!45, line width=0.7pt, rounded corners=3pt, fill=cPrep!4,
                   minimum height=0.88cm, minimum width=14.46cm, anchor=west},
  badge/.style  = {fill=cPrep, text=white, rounded corners=1.8pt, font=\scriptsize\bfseries,
                   inner xsep=4pt, inner ysep=2pt, anchor=west},
  badge1/.style = {fill=black!55, text=white, rounded corners=1.8pt, font=\scriptsize\bfseries,
                   inner xsep=4pt, inner ysep=2pt, anchor=east},
  nm/.style     = {font=\footnotesize, anchor=west},
  chip/.style   = {draw=cPrep!45, fill=white, rounded corners=2pt, font=\scriptsize,
                   inner ysep=3pt, minimum height=0.42cm, anchor=west, align=center},
  chipon/.style = {chip, draw=cPrep!70, fill=cPrep!14},
  minichip/.style={draw=cData!55, fill=white, rounded corners=2pt, font=\scriptsize,
                   inner ysep=2.5pt, minimum height=0.38cm, minimum width=1.40cm,
                   anchor=center, text=black!70},
  orx/.style    = {font=\scriptsize\itshape, black!45, anchor=west},
  note/.style   = {font=\footnotesize\itshape, black!55},
]

\node[stage=cPrep] (P) at ( 6.00,0) {};
\node[stage=cTune] (T) at (10.52,0) {};
\node[stage=cDec]  (D) at (15.04,0) {};
\node[stage=cEval] (E) at (19.56,0) {};
\foreach \a/\b in {P/T, T/D, D/E} \draw[arr] (\a.east) -- (\b.west);

\node[inbox] (dset) at (3.00,-1.60) {Dataset};
\node[inbox] (sens) at (3.00,-3.35) {Sensitive Attributes};

\node[font=\scriptsize, black!52, anchor=center] at (3.00,-2.16)
  {MovieLens-1M $\cdot$ 3,883 items};
\begin{scope}[shift={(2.30,-2.66)}, line width=0.6pt]
  \begin{scope}                                              
    \draw[cData, fill=cData!12] (-0.15,-0.11) rectangle (0.15,0.11);
    \draw[cData] (-0.09,-0.11) -- (-0.09,0.11); \draw[cData] (0.09,-0.11) -- (0.09,0.11);
    \foreach \y in {0.065,-0.065}{\foreach \x in {-0.12,0.12}{
      \fill[cData!65] (\x-0.024,\y-0.021) rectangle (\x+0.024,\y+0.021);}}
  \end{scope}
  \begin{scope}[shift={(0.47,0)}]                            
    \draw[cData, fill=cData!12] (-0.15,-0.11) rectangle (0.15,0.04);
    \draw[cData, fill=cData!45] (-0.15,0.04) rectangle (0.15,0.12);
    \foreach \x in {-0.11,-0.03,0.05}
      \draw[white, line width=0.55pt] (\x,0.04) -- (\x+0.045,0.12);
  \end{scope}
  \begin{scope}[shift={(0.94,0)}]                            
    \draw[cData, fill=cData!12, rounded corners=1pt] (-0.13,-0.13) rectangle (0.13,0.13);
    \fill[cData] (-0.045,-0.075) -- (-0.045,0.075) -- (0.085,0) -- cycle;
  \end{scope}
  \begin{scope}[shift={(1.40,0)}]                            
    \fill[cData!70] (90:0.145) -- (126:0.060) -- (162:0.145) -- (198:0.060) -- (234:0.145)
                 -- (270:0.060) -- (306:0.145) -- (342:0.060) -- (18:0.145) -- (54:0.060) -- cycle;
  \end{scope}
\end{scope}

\node[minichip] at (2.23,-3.93) {Gender};
\node[minichip] at (3.77,-3.93) {Age};

\draw[lead] (dset.east) -- (5.00,-1.60);
\draw[lead] (sens.east) -- (5.00,-3.35);
\draw[lead, -{Latex[length=2.2mm,width=1.6mm]}] (5.00,-3.35) -- (5.00,-0.90);

\begin{scope}[shift={(4.45,0)}, draw=cPrep, line width=0.85pt]         
  \foreach \y/\x in {0.155/0.12, 0/-0.09, -0.155/0.19}{
    \draw (-0.30,\y) -- (0.30,\y); \fill[cPrep] (\x,\y) circle (0.062);}
\end{scope}
\begin{scope}[shift={(8.97,0)}, draw=cTune, line width=0.85pt]         
  \draw[rounded corners=1.5pt, fill=cTune!12] (-0.21,-0.21) rectangle (0.21,0.21);
  \draw[fill=white] (-0.095,-0.095) rectangle (0.095,0.095);
  \foreach \d in {-0.12,0,0.12}{
    \draw (-0.21,\d) -- (-0.32,\d); \draw (0.21,\d) -- (0.32,\d);
    \draw (\d,0.21) -- (\d,0.32);   \draw (\d,-0.21) -- (\d,-0.32);}
\end{scope}
\begin{scope}[shift={(13.49,0.24)}, line width=0.8pt]                  
  \draw[cDec] (0,0) -- (-0.22,-0.24);  \draw[cDec] (0,0) -- (0.08,-0.24);
  \draw[black!30, dash pattern=on 1.2pt off 1.2pt] (0,0) -- (0.23,-0.21);
  \draw[cDec] (-0.22,-0.24) -- (-0.22,-0.48);
  \fill[cDec] (0,0) circle (0.064);
  \fill[cDec] (-0.22,-0.24) circle (0.055);  \fill[cDec] (0.08,-0.24) circle (0.055);
  \fill[cDec] (-0.22,-0.48) circle (0.055);
  \draw[black!30, fill=white] (0.23,-0.21) circle (0.051);
\end{scope}
\begin{scope}[shift={(18.01,0.08)}, line width=0.7pt]                  
  \foreach \i in {0,1,2}{\fill[cEval!45] (-0.30,-0.10*\i) rectangle (-0.02,-0.10*\i+0.056);}
  \foreach \i in {0,1}{  \fill[cEval!45] ( 0.06,-0.10*\i) rectangle ( 0.34,-0.10*\i+0.056);}
  \fill[cPrep!60] (0.06,-0.200) rectangle (0.34,-0.144);
\end{scope}

\node[ftxt] at (4.90,0) {\renewcommand{\arraystretch}{1.14}\begin{tabular}{@{}l@{}}
  {\small\bfseries\textcolor{cPrep}{Preprocessing Choices}}\\
  {\small\bfseries\textcolor{cPrep}{in Fine-Tuning Data}}\\[1.5pt]
  {\color{black!60}Three Independent}\\
  {\color{black!60}Choices}
\end{tabular}};
\node[ftxt] at (9.42,0) {\renewcommand{\arraystretch}{1.14}\begin{tabular}{@{}l@{}}
  {\small\bfseries\textcolor{cTune}{Controlled Fine-Tuning}}\\[1.5pt]
  {\color{black!60}Prompt Augmentation}\\
  {\color{black!60}Mixed Fine-Tuning Corpus}\\
  {\color{black!60}Output Contract}
\end{tabular}};
\node[ftxt] at (13.94,0) {\renewcommand{\arraystretch}{1.14}\begin{tabular}{@{}l@{}}
  {\small\bfseries\textcolor{cDec}{Catalog-Constrained}}\\
  {\small\bfseries\textcolor{cDec}{Recommendation}}\\[1.5pt]
  {\color{black!60}Prefix Tree over}\\
  {\color{black!60}Item Titles}
\end{tabular}};
\node[ftxt] at (18.46,0) {\renewcommand{\arraystretch}{1.14}\begin{tabular}{@{}l@{}}
  {\small\bfseries\textcolor{cEval}{Controlled Evaluation}}\\[1.5pt]
  {\color{black!60}Utility}\\
  {\color{black!60}Output Control}\\
  {\color{black!60}Demographic Sensitivity}
\end{tabular}};

\draw[black!42, line width=0.7pt] (8.52,1.06) -- (8.52,1.22) -- (21.56,1.22) -- (21.56,1.06);
\node[badge1] at (12.15,1.44) {RQ1};
\node[note, anchor=west] at (12.30,1.44) {Controlled backbone, identical in every comparison};

\node[note, anchor=west] at (6.75,-1.15) {One choice varies at a time};
\draw[cPrep!65, line width=0.85pt] (6.60,-0.90) -- (6.60,-3.80);
\foreach \y in {-1.80,-2.80,-3.80} \draw[cPrep!65, line width=0.85pt] (6.60,\y) -- (7.10,\y);

\node[rqrow] at (7.10,-1.80) {};  \node[badge] at (7.30,-1.80) {RQ2};
\node[nm] at (8.75,-1.80) {Item-Title Representation};
\node[chip,   minimum width=3.60cm] at (13.30,-1.80) {\texttt{The Matrix (1999)}};
\node[orx] at (17.05,-1.80) {or};
\node[chipon, minimum width=3.60cm] at (17.45,-1.80) {\texttt{Matrix, The (1999)}};

\node[rqrow] at (7.10,-2.80) {};  \node[badge] at (7.30,-2.80) {RQ3};
\node[nm] at (8.75,-2.80) {Demographic Balancing};
\node[chip,   minimum width=2.25cm] at (13.30,-2.80) {As Observed};
\node[orx] at (15.70,-2.80) {or};
\node[chipon, minimum width=2.25cm] at (16.10,-2.80) {Gender-Balanced};
\node[orx] at (18.50,-2.80) {or};
\node[chipon, minimum width=2.25cm] at (18.90,-2.80) {Age-Balanced};

\node[rqrow] at (7.10,-3.80) {};  \node[badge] at (7.30,-3.80) {RQ4};
\node[nm] at (8.75,-3.80) {Knowledge Enrichment};
\node[chip,   minimum width=3.60cm] at (13.30,-3.80) {Recommendation Only};
\node[orx] at (17.05,-3.80) {or};
\node[chipon, minimum width=3.60cm] at (17.45,-3.80) {+ Genre and Plot QA};

\begin{scope}[shift={(8.35,-1.80)}]                                    
  \draw[cPrep!70, fill=cPrep!10, rounded corners=1pt, line width=0.6pt]
    (-0.17,-0.115) rectangle (0.17,0.115);
  \draw[cPrep!70, line width=0.5pt] (-0.105,0.038) -- (0.105,0.038);
  \draw[cPrep!70, line width=0.5pt] (-0.105,-0.038) -- (0.040,-0.038);
\end{scope}
\begin{scope}[shift={(8.35,-2.80)}]                                    
  \fill[cPrep!70] (-0.095,0.050) circle (0.050);
  \fill[cPrep!70] (-0.160,-0.120) rectangle (-0.030,-0.005);
  \draw[cPrep!70, line width=0.6pt] (0.095,0.050) circle (0.050);
  \draw[cPrep!70, line width=0.6pt] (0.030,-0.120) rectangle (0.160,-0.005);
\end{scope}
\begin{scope}[shift={(8.35,-3.80)}]                                    
  \draw[cPrep!70, fill=white, rounded corners=1pt, line width=0.6pt]
    (-0.135,-0.130) rectangle (0.045,0.130);
  \draw[cPrep!70, line width=0.5pt] (-0.085,0.062) -- (0.000,0.062);
  \draw[cPrep!70, line width=0.5pt] (-0.085,0.000) -- (0.000,0.000);
  \draw[cPrep!70, line width=0.7pt] (0.090,-0.060) -- (0.190,-0.060);
  \draw[cPrep!70, line width=0.7pt] (0.140,-0.110) -- (0.140,-0.010);
\end{scope}

\end{tikzpicture}}
\caption{Overview of the controlled experimental pipeline.}
\Description{A left-to-right overview of the data preparation, preprocessing
choices, fine-tuning, catalog-constrained recommendation generation, and
controlled evaluation pipeline.}
\label{fig:pipeline-overview}
\end{figure*}

Figure~\ref{fig:pipeline-overview} presents an overview of the controlled experimental pipeline used in this study. We first establish a shared catalog-constrained recommendation backbone and then study three preprocessing choices in fine-tuning data: item-title representation, demographic balancing, and auxiliary item-knowledge enrichment. We then evaluate recommendation utility, output behavior, prompt consistency, and demographic sensitivity.

\subsection{Task Formulation}
Let $\mathcal{U}$ denote a set of users and $\mathcal{I}$ a fixed item catalog. For each user $u \in \mathcal{U}$, we observe a chronologically ordered interaction history $H_u = (i_{u,1}, i_{u,2}, \ldots, i_{u,n_u})$, where $n_u$ is the history length of user $u$ and each $i_{u,j} \in \mathcal{I}$ is an item associated with a positive preference signal. The task is to generate an ordered recommendation list \[ R_u = (r_{u,1}, r_{u,2}, \ldots, r_{u,K}) \] conditioned on $H_u$, where each $r_{u,j} \in \mathcal{I}$ and the recommended items are expected to be distinct.

We formulate recommendation as conditional language generation. A prompt function $\phi$ converts a user history and an optional descriptor $d$ into a textual input $x_u$, with $d=\varnothing$ denoting a neutral prompt. The generation process $f_\theta$, comprising the fine-tuned model and its decoding procedure, produces a token sequence $\widehat{y}_u$, which is parsed into a catalog-item recommendation list $\widehat{R}_u$ by a representation-specific parser $\pi_\rho$:
\[
x_u=\phi(H_u;d), \qquad
\widehat{y}_u=f_\theta(x_u;\xi_u), \qquad
\widehat{R}_u=\pi_\rho(\widehat{y}_u).
\]
Here, $\xi_u$ denotes decoding randomness when stochastic decoding is used, and $\rho$ denotes the item-representation configuration; for readability, the dependence of the decoding procedure on $\rho$ is suppressed in the notation. Within $\widehat{y}_u$, each item span is delimited by \texttt{<SOI>} and \texttt{<EOI>} and represents a single serialized catalog item. Recommendation metrics are computed on $\widehat{R}_u$, while structural deviations such as repeated items are retained as observable output behavior and evaluated separately.


Each supervised recommendation example consists of an input prompt $x_u$ and a serialized target sequence
$y_u=(y_{u,1},\ldots,y_{u,L_u})$, where $L_u$ denotes its token length.


An item-representation configuration $\rho$ defines a serialization function $s_\rho$ that determines how catalog items are represented in the textual recommendation pipeline, including training targets, parser mappings, and decoding constraints. Let $\tau_\rho(i)$ denote the token sequence obtained by tokenizing $s_\rho(i)$ under the tokenizer and formatting convention used within an item span. Changing $s_\rho$ leaves the underlying catalog unchanged but may alter tokenization, prefix sharing, and the set of token sequences the language model can generate for catalog items. To ensure that $\pi_\rho$ maps generated item spans back to catalog items unambiguously, serialized representations must uniquely identify items in the catalog under evaluation; representation-induced collisions are therefore resolved before training and evaluation.

We study three preprocessing choices in fine-tuning data: item representation, demographic balancing, and auxiliary item knowledge. Demographic balancing modifies the sampling distribution of fine-tuning examples without introducing a fairness-specific objective, whereas auxiliary item knowledge adds supplementary item-information examples to the fine-tuning corpus.


\subsection{Training Objective and Output Contract}

The model is trained to generate recommendation lists using explicit item-span markers. Each catalog item in the target sequence is serialized as
\[
\texttt{\textless SOI\textgreater}\;s_\rho(i)\;\texttt{\textless EOI\textgreater}.
\]
The start-of-item (SOI) marker indicates the beginning of a catalog item span, while the end-of-item (EOI) marker closes the span. This output contract provides a deterministic parsing rule for well-formed outputs and defines the interface between language generation and catalog-con\-strained decoding.

The model is trained using scope-masked cross-entropy. Outside item spans, the objective is equivalent to ordinary autoregressive cross-entropy, whereas within item spans the candidate vocabulary is restricted to catalog-valid continuations when multiple valid continuations are available. Let
\[
\mathcal{C}_\rho=\{\tau_\rho(i):i\in\mathcal{I}\}
\]
denote the tokenized catalog, and let $\mathcal{V}$ denote the model token vocabulary. For a target position $\ell$ inside an item span, let $q_{u,\ell}$ denote the target item-token prefix following the most recent \texttt{\textless SOI\textgreater} and preceding position $\ell$. The set of valid next tokens for a prefix $q$ is
\[
A_\rho(q)
=
\{v\in\mathcal{V} :
\exists s\in\mathcal{C}_\rho,\;
q\mathbin{\|}(v)\preceq s\}
\;\cup\;
\{\texttt{\textless EOI\textgreater}:q\in\mathcal{C}_\rho\},
\]
where $\mathbin{\|}$ denotes sequence concatenation and $\preceq$ denotes the sequence-prefix relation. The closing marker is therefore available only when the current prefix corresponds to a complete catalog item, including cases in which that item is also a prefix of a longer catalog representation.

For the purpose of defining the candidate vocabulary, the closing marker is treated as part of the item span. The candidate vocabulary at target position $\ell$ is then defined as
\[
S_{u,\ell}
=
\begin{cases}
A_\rho(q_{u,\ell}),
& \text{if $\ell$ is within an item span and } |A_\rho(q_{u,\ell})|>1,\\
\mathcal{V},
& \text{otherwise}.
\end{cases}
\]

The resulting training objective is
\[
\mathcal{L}_{\mathrm{scope}}(u)
=
-\sum_{\ell=1}^{L_u}
\log
\frac{
p_\theta(y_{u,\ell}\mid x_u,y_{u,<\ell})
}{
\sum_{v\in S_{u,\ell}}
p_\theta(v\mid x_u,y_{u,<\ell})
}.
\]

When $S_{u,\ell}=\mathcal{V}$, the denominator equals one and the objective reduces to standard autoregressive cross-entropy. At positions within item spans where multiple catalog-valid continuations are available, the loss is normalized over those valid continuations. If an item prefix has only one valid continuation, the full vocabulary is retained to avoid a one-candidate softmax that would otherwise contribute zero training loss.

\subsection{Catalog-Constrained Generation}

Unconstrained LLM-based recommenders may generate plausible items that are not part of the target catalog. Following recent work on eliminating out-of-catalog recommendations in LLM-based recommender systems \cite{liao2026eliminating}, we use prefix-tree constrained decoding to ground item generation in the fixed catalog.

For a given item representation $\rho$, the token sequences in $\mathcal{C}_\rho$ are inserted into a prefix tree, where each completed path corresponds to a serialized catalog item. When the model emits \texttt{\textless SOI\textgreater}, decoding enters catalog-con\-strained mode. At generation position $\ell$, let $\widehat{q}_{u,\ell}$ denote the generated within-item prefix following the most recent \texttt{\textless SOI\textgreater} and preceding position $\ell$. The next-token distribution is restricted to the valid continuation set
$A_\rho(\widehat{q}_{u,\ell})$:
\[
\widetilde{p}_{\theta}
\left(
v \mid x_u,\widehat{y}_{u,<\ell}
\right)
=
\frac{
p_{\theta}
\left(
v \mid x_u,\widehat{y}_{u,<\ell}
\right)
\mathbf{1}[v\in A_\rho(\widehat{q}_{u,\ell})]
}{
\sum_{w\in A_\rho(\widehat{q}_{u,\ell})}
p_{\theta}
\left(
w \mid x_u,\widehat{y}_{u,<\ell}
\right)
}.
\]

The \texttt{\textless EOI\textgreater} marker becomes available only when the current prefix corresponds to a complete catalog-item sequence. Emitting this marker closes the item span and returns decoding to the ordinary generation mode. Outside item spans, generation proceeds over the full model vocabulary.

Catalog-constrained decoding removes out-of-catalog generation as a confounding factor by preventing invalid catalog strings within item spans, but it does not guarantee recommendation relevance, diversity, or fairness.

\begin{figure}[t]
  \centering
  \begin{tikzpicture}[
    grow=down,
    level distance=9mm,
    level 1/.style={sibling distance=24mm},
    level 2/.style={sibling distance=16mm},
    level 3/.style={sibling distance=12mm},
    edge from parent/.style={draw=green!45!black, thick, -{Latex[length=1.5mm]}},
    root/.style={circle, draw=cyan!55!black, fill=cyan!8, thick,
      font=\scriptsize\ttfamily, inner sep=2.5pt},
    allowed/.style={rounded corners=2pt, draw=green!45!black, fill=green!7,
      thick, align=center, font=\scriptsize\ttfamily,
      inner xsep=4pt, inner ysep=2.5pt},
    completion/.style={rounded corners=2pt, draw=cyan!55!black, fill=cyan!8,
      thick, align=center, font=\scriptsize\ttfamily,
      inner xsep=4pt, inner ysep=2.5pt},
    terminal/.style={circle, draw=cyan!55!black, fill=cyan!8, thick,
      font=\tiny\ttfamily, inner sep=2pt},
    masked/.style={rounded corners=2pt, draw=red!65!black, fill=red!5,
      dashed, thick, align=center, text=red!65!black,
      font=\scriptsize\ttfamily, inner xsep=4pt, inner ysep=2.5pt}
  ]
    \node[root] {\textless SOI\textgreater}
      child { node[allowed] {$v_1$\\[-1pt]\tiny allowed}
        child { node[completion] {$v_3$\\[-1pt]\tiny completes $s_\rho(i_a)$}
          child { node[terminal] {\textless EOI\textgreater} }
        }
      }
      child { node[allowed] {$v_2$\\[-1pt]\tiny allowed}
        child { node[completion] {$v_4$\\[-1pt]\tiny completes $s_\rho(i_b)$}
          child { node[terminal] {\textless EOI\textgreater} }
        }
      }
      child[edge from parent/.style={draw=red!65!black, dashed, thick,
        -{Latex[length=1.5mm]}}] { node[masked] {$v_x$\\[-1pt]\tiny masked} };
  \end{tikzpicture}
  \caption{Prefix-tree constrained decoding within an item span. After the model emits \texttt{\textless SOI\textgreater}, only tokens that continue a valid catalog path remain available. Invalid continuations are masked, and \texttt{\textless EOI\textgreater} becomes available only after a complete catalog-item sequence has been generated.}
  \Description{A prefix tree rooted at the start-of-item token. Two allowed branches lead to valid catalog-item sequences and end-of-item nodes, while one invalid branch is masked.}
  \label{fig:prefix-tree-example}
\end{figure}
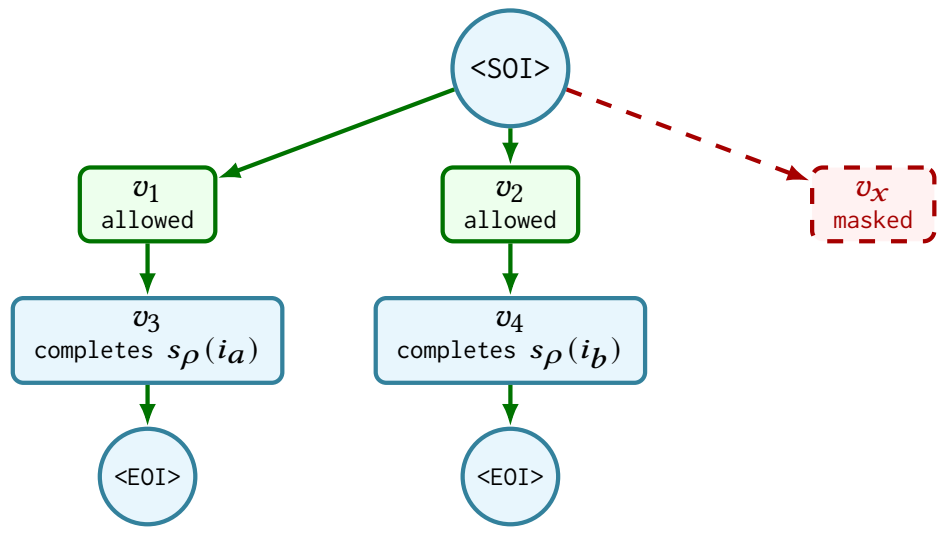

Because the prefix tree is constructed from the representation-specific catalog sequences $\mathcal{C}_\rho$, item representation determines both the training target space and the constrained generation space. Figure~\ref{fig:prefix-tree-example} illustrates how valid and invalid continuations are handled within an item span.

The decoder does not explicitly enforce exactly $K$ completed item spans, prevent early sequence termination, or mask previously generated items. Therefore, Exact-$K$ compliance and repetition are treated as empirical output-control measures rather than guaranteed properties of constrained decoding.

\subsection{Prompt Augmentation}

Language model behavior can change across instruction paraphrases, prompt templates, and small textual perturbations, even when the task itself is unchanged \cite{mizrahi-etal-2024-state,chatterjee-etal-2024-posix}. A model fine-tuned on a single instruction wording may therefore become dependent on that particular presentation. To reduce this dependence, for prompt-augmented configurations, we expose the model during fine-tuning to multiple prompt formulations of the recommendation instruction. The goal is to encourage robustness to changes in prompt wording and presentation format while preserving the underlying recommendation task.

Prompt Perturbation Consistency Learning (PPCL) addresses prompt sensitivity by introducing a consistency objective between clean and perturbed prompts \cite{qiang-etal-2024-prompt}. Inspired by this idea, we use template-based prompt augmentation, but incorporate the augmented prompts directly into fine-tuning without an additional consistency objective.

Formally, we extend the prompt function introduced in the task formulation from a single function $\phi$ to a collection of prompt templates $ \Phi=\{\phi_1,\phi_2,\ldots,\phi_M\}$.
For a fixed user history $H_u$ and descriptor $d$, template $\phi_m\in\Phi$ constructs the input \[ x_u^{(m)}=\phi_m(H_u;d). \] During prompt-augmented fine-tuning, we set $d=\varnothing$.
The templates vary the instruction wording, constraint emphasis, preference framing, prompt format and length, and minor text perturbations, while preserving the underlying interaction history, serialized target sequence, item representation, and recommendation objective.

In addition to prompt instruction variation, we also vary how the interaction history $H_u$ is presented in the prompt without changing its item content or ordering. 
During construction of a supervised recommendation example, a prompt template and one of the supported history formats are selected to form the input. For a fixed example and item-representation configuration $\rho$, the target sequence $y_u$, catalog $\mathcal{I}$, serialization function $s_\rho$, and output contract remain unchanged. Thus, prompt augmentation changes only how the input is expressed, while preserving the recommendation task and its target space.

In contrast to training-time prompt augmentation, attribute-conditioned prompting is used during evaluation to measure sensitivity to demographic descriptors while keeping the underlying recommendation context fixed.

\subsection{Attribute-Conditioned Prompt Evaluation}

To evaluate fairness-related behavior, we compare recommendations generated from the same interaction history under neutral and attribute-conditioned prompt settings. Let $a$ denote a user attribute and let $g \in \mathcal{G}_a$ denote one of its possible values or groups. For a fixed evaluation prompt function $\phi_{\mathrm{eval}}$ and user history $H_u$, the neutral and attribute-conditioned prompts are defined as
\[
x_u^0 = \phi_{\mathrm{eval}}(H_u;\varnothing),
\qquad
x_u^{a,g} = \phi_{\mathrm{eval}}(H_u;(a,g)).
\]
The interaction history, recommendation instruction, prompt template, decoding configuration, and target catalog are held fixed across the paired conditions; only the injected attribute descriptor changes. Under stochastic decoding, the same decoding randomness $\xi_u$ is used for both conditions to reduce variation unrelated to the attribute intervention.
The corresponding recommendation lists are
\[
\widehat{R}_u^0
=
\pi_\rho
\left(
f_\theta(x_u^0;\xi_u)
\right),
\qquad
\widehat{R}_u^{a,g}
=
\pi_\rho
\left(
f_\theta(x_u^{a,g};\xi_u)
\right).
\]

Differences between $\widehat{R}_u^0$ and $\widehat{R}_u^{a,g}$ are treated as attribute-sensitivity signals, since only the attribute descriptor varies across the paired conditions. These comparisons are aggregated across users for each evaluated attribute and group.

Each pair yields a set overlap between $\widehat{R}_u^0$ and $\widehat{R}_u^{a,g}$, which we report as a Jaccard similarity and average within each group $g$. The spread of these group averages across $\mathcal{G}_a$ is reported as SNSR, and is small when the descriptor shifts every group by a similar amount. Group-level utility differences are measured separately under the neutral prompt and reported as the Neutral Gap. Section~\ref{sec:metrics} gives the definitions.

This protocol measures fairness-related behavioral variation rather than fairness in an absolute sense. Such differences do not by themselves constitute evidence of discrimination and are therefore interpreted together with recommendation utility and other behavioral measures.

\section{Experimental Setup}

This section describes the dataset, chronological split, and preparation of fine-tuning data, followed by the implementation details and evaluation metrics used across the experiments.

\subsection{Dataset, Chronological Split, and Fine-Tuning Data Preparation}

We conduct our experiments on MovieLens-1M, which contains 6,040 users, 3,883 movies, and 1,000,209 ratings \cite{harper2015movielens}. Ratings greater than or equal to 4 are treated as positive interactions. Users with fewer than 22 positive interactions are excluded, ensuring sufficient interactions for the maximum-length training instance of ten history items and ten targets while reserving one interaction each for validation and testing. These criteria retain 4,995 users. For each user, positive interactions are ordered chronologically; the final interaction is separated as the test target, the preceding interaction is used for validation, and the remaining earlier interactions are used for the training set. This split yields one validation target and one test target for each of the 4,995 retained users. The recommendation fine-tuning records are then sampled from the remaining training interactions, as described below. The full catalog of 3,883 movies is retained for constrained generation; 3,432 items (88.4\%) occur in the training interactions, while 451 items (11.6\%) remain unseen in the recommendation fine-tuning data.

Recommendation fine-tuning examples are constructed as history-to-target generation instances. For each user, we sample 20 training instances from the chronological training sequence, yielding 99,900 recommendation examples in total. Each instance contains a variable-length history and an ordered target list, with both lengths capped at ten items. The requested target count $K$ is written into the instruction, so the prompt and the target sequence always agree on list length. Extremely long raw titles are filtered using a 99th-percentile title-length cutoff to avoid formatting outliers. Histories are serialized using comma-separated, arrow-separated, semicolon-separated, and numbered-list formats while preserving chronological item order.

We evaluate two item-title representations. The normalized-title condition moves trailing articles to the beginning of the title, e.g., \textit{Matrix, The (1999)} becomes \textit{The Matrix (1999)}. The original-title condition preserves the MovieLens surface form, e.g., \textit{Matrix, The (1999)}. This distinction affects the token-level structure of the catalog: normalization rewrites 714 of the 3,883 titles, and the resulting catalog places 715 titles, or 18.4\% of the catalog, behind one shared opening token. Under the original MovieLens form, no opening token accounts for more than 1.8\% of the catalog. Both representations are applied consistently to training targets, parsing, and prefix-tree constrained decoding.

Following the RecLM-cgen design, which augments recommendation fine-tuning with conversational data to help preserve general conversational ability \cite{liao2026eliminating}, we construct a mixed fine-tuning corpus containing 119,056 examples: 99,900 recommendation examples (83.9\%), 14,161 general conversational examples (11.9\%), and 4,995 recommendation-oriented conversational examples (4.2\%). These proportions are design choices guided by prior mixed fine-tuning practice rather than fixed values copied from prior work. 

To reduce dependence on a single prompt wording, recommendation examples are expressed using 37 semantically equivalent instruction templates organized into six families as given in Table~\ref{tab:prompt-families}. The templates differ in phrasing rather than task semantics, spanning direct recommendation, explicit rule lists, preference framing, output-format emphasis, compressed phrasing, and minor misspellings. The same prompt-augmentation strategy is applied across the compared fine-tuning configurations. For example, a T6 template replaces \textit{Recommend} with \textit{Recomend} while keeping the history, requested list length, and output requirements unchanged.

\begin{table}[!t]
\centering
\caption{Prompt template families used during fine-tuning. Shares give the proportion of recommendation examples assigned to each family.}
\label{tab:prompt-families}
\small
\begin{tabular}{llrr}
\toprule
Family & Character & Templates & Share \\
\midrule
T1 & Direct request & 8 & 30\% \\
T2 & Explicit rule list & 7 & 25\% \\
T3 & Preference framing & 7 & 20\% \\
T4 & Output format emphasis & 6 & 15\% \\
T5 & Compressed phrasing & 5 & 7\% \\
T6 & Misspelled instructions & 4 & 3\% \\
\midrule
Total & & 37 & 100\% \\
\bottomrule
\end{tabular}
\end{table}

For the demographic-balancing conditions used in RQ3, we modify only the recommendation fine-tuning examples. The recommendation corpus size is held fixed at 99,900 examples, and the two groups of the balanced attribute receive 49,950 examples each. Under gender balancing, 3,620 male users contribute 13.8 examples each on average and 1,375 female users contribute 36.3. Under age balancing, using the binary split at 35, 2,152 older users contribute 23.2 examples each and 2,843 younger users contribute 17.6. Demographic labels are used only for constructing these balanced corpora and are not included in the recommendation prompts.

For the knowledge-enriched condition used in RQ4, we add 7,683 auxiliary item-level question--answer examples about catalog movies. These examples cover genre and plot-style knowledge: genres are taken from MovieLens metadata, and plot summaries are taken from TMDB and joined to the catalog by item identifier. Answers that name a movie use the same \texttt{<SOI>} and \texttt{<EOI>} item-span markers as the recommendation task. The knowledge examples are added to the fine-tuning corpus rather than replacing recommendation examples, resulting in 126,739 total fine-tuning examples.

\subsection{Implementation Details}


All configurations fine-tune Llama-3-8B-Instruct \cite{llama2024herd}. The base weights are loaded in 4-bit NF4 with double quantization and a bfloat16 compute type, and low-rank adapters are trained on top, following QLoRA \cite{dettmers2023qlora,hu2022lora}. LoRA adapters are applied to the attention query, key, value, and output projections and to the feed-forward gate, up, and down projections. We use rank $r=16$, scaling factor $\alpha=32$, and dropout 0.05. Because \texttt{<SOI>} and \texttt{<EOI>} are added as new tokenizer entries, the input embedding matrix and output language-model head are trained alongside the adapters.

Optimization uses paged 8-bit AdamW with a learning rate of $10^{-4}$, a cosine schedule, and a warmup ratio of 0.03. Weight decay is set to zero and gradients are clipped at a norm of 1.0. Attention is computed with FlashAttention-2, and training uses a device batch of one sequence and gradient accumulation over 16 steps, giving an effective batch size of 16. Training performs a single pass over the fine-tuning corpus, so each example is seen exactly once.

At inference time, models are served with vLLM \cite{kwon2023vllm}. All headline evaluations use greedy decoding at temperature zero, a context window of 1,024 tokens, and a generation budget of 512 new tokens. Item spans are decoded with the prefix-tree constraint unless an unconstrained ablation is explicitly reported. Each test user is evaluated with the same ten-item history and a request for ten recommendations.
Each configuration is based on a single fine-tuning run with fixed data-generation and training seeds and is evaluated deterministically using a fixed checkpoint and prompt set.

\subsection{Evaluation Metrics}\label{sec:metrics}

We report metrics covering five complementary dimensions: recommendation utility, catalog usage, output control, prompt consistency, and demographic sensitivity. Recommendation utility is measured with Hit Ratio@10 (HR@10), which indicates whether the held-out test item appears among the ten generated recommendations. Catalog usage is summarized with Coverage@10 and Unique Items. Coverage@10 is the proportion of catalog items that appear at least once across all generated top-10 lists, while Unique Items reports the corresponding count of distinct recommended catalog items.

Output control is evaluated with Exact-10, Out of Domain@10 (OOD@10), and Repeat@10. Exact-10 is the proportion of responses containing exactly ten extracted item spans. OOD@10 measures the proportion of generated recommendations that do not correspond to valid catalog items. Repeat@10 measures within-list duplication among the generated recommendations. Higher Exact-10 and lower OOD@10 and Repeat@10 indicate more reliable recommendation outputs.

Prompt consistency is evaluated in RQ1 using a separate set of five evaluation prompt variants. The variants preserve the same user history, target list length, and output contract while changing the request wording through a clean reference, paraphrase, instruction-wording change, format edit, and compact prompt. Prompt HR Spread is the difference between the highest and lowest HR@10 across the five variants. Prompt Jaccard compares each non-reference variant with the clean reference prompt for the same users and averages the resulting top-10 set overlaps across users and variants.

Fairness-related behavior is evaluated through paired prompt comparisons. For each user, we compare recommendations generated from the same interaction history under a neutral prompt and a demographic-conditioned prompt. Neutral-conditioned Jaccard measures the overlap between these two top-10 recommendation sets; higher values indicate that adding a demographic descriptor changes the recommendation list less. Pair Jaccard, used in RQ3, directly compares recommendation sets generated for the same history under two demographic descriptor values, such as male versus female or age $<35$ versus age $\geq35$.

We also report Neutral Gap and Sensitive-to-Neutral Similarity Range (SNSR) \cite{zhang2023chatgptfair}. Neutral Gap is the absolute difference in HR@10 between demographic groups under the neutral prompt, reported in percentage points. For a demographic attribute with groups $\mathcal{G}_a$, let $\overline{J}_g$ denote the average neutral-conditioned Jaccard similarity for users in group $g$. SNSR is defined as
\[
\mathrm{SNSR}
=
\max_{g \in \mathcal{G}_a} \overline{J}_g
-
\min_{g \in \mathcal{G}_a} \overline{J}_g .
\]
Lower SNSR indicates that the demographic descriptor affects the evaluated groups more uniformly. We interpret Neutral Gap, Neutral-conditioned Jaccard, Pair Jaccard, and SNSR as indicators of demographic sensitivity and group-level utility disparity, not as direct evidence of discrimination.

\section{Experimental Results}
We organize the experimental analysis around four research questions that examine the reliability of the recommendation pipeline and the effects of preprocessing choices in fine-tuning data. We first establish the controlled recommendation backbone and then evaluate item-title representation, demographic balancing, and additional item information.
\subsection{RQ1: Does the recommendation pipeline provide sufficient catalog validity, output control, and prompt consistency for controlled comparisons of preprocessing choices in fine-tuning data?}
Generative recommenders can return lists of the wrong length, repeat items, or name items that are not in the catalog \cite{munson2025review,liao2026eliminating}. RQ1 therefore examines whether the pipeline provides a sufficiently controlled basis for the later comparisons of preprocessing choices and demographic sensitivity. Recommendation utility is part of that basis. A model whose lists are only weakly tied to the input can appear insensitive to demographic descriptors simply because it recommends poorly, so stability under demographic prompting is informative only once the pipeline recovers held-out preferences at a reasonable rate.
We first evaluate a structured supervised fine-tuning (SFT) baseline under unconstrained decoding and then apply prefix-tree constrained decoding to the same SFT weights to isolate the effect of catalog grounding. We next introduce the catalog-con\-strained training configuration and finally add prompt augmentation while keeping the catalog-con\-strained generation mechanism fixed. Prompt consistency is evaluated separately using five evaluation prompt variants. Table~\ref{tab:rq1-reliability} summarizes these configurations.

The SFT baseline learns the structured \texttt{\textless SOI\textgreater}--\texttt{\textless EOI\textgreater} output format, but SFT alone does not provide reliable list-level control. Although it achieves an HR@10 of 7.47\%, only 0.06\% of its outputs contain exactly ten recommendations, and 2.08\% of its recommended items fall outside the catalog. This indicates that learning the serialization format alone does not prevent variable-length or out-of-catalog outputs.

Applying the prefix tree to the same SFT weights (SFT constrained decoding) reduces OOD@10 from 2.08\% to 0.00\% while leaving HR@10 essentially unchanged (7.47\% versus 7.43\%). This comparison isolates the effect of catalog grounding and shows that constrained decoding can eliminate invalid catalog items without substantially changing recommendation utility. However, constrained decoding alone does not provide list-level control: Exact-10 remains near zero (0.02\%), while Repeat@10 increases from 3.12\% to 4.86\%.

The catalog-con\-strained training configuration improves list-level consistency. Exact-10 increases to 100\%, OOD@10 remains at 0.00\%, and HR@10 improves to 9.89\%. This configuration produces consistently sized recommendation lists while restricting generated item spans to catalog-valid paths, thereby providing a more controlled output space for subsequent comparisons. However, repeated-item behavior remains substantial, with Repeat@10 reaching 10.56\%. Thus, catalog validity and exact-length compliance do not by themselves eliminate within-list redundancy.

The prompt-augmented model retains the same catalog-con\-strained decoding mechanism while exposing the model to multiple semantically equivalent recommendation instructions during fine-tuning. The resulting model preserves 100\% Exact-10 compliance and zero out-of-catalog generation, while improving HR@10 from 9.89\% to 10.67\% and reducing Repeat@10 from 10.56\% to 8.01\%.
Prompt augmentation also reduces sensitivity to prompt formulation: Prompt HR Spread decreases from 2.18 to 0.26 percentage points, while Prompt Jaccard increases from 0.6799 to 0.7489. These results indicate that prompt-augmented fine-tuning improves recommendation utility and prompt-level consistency while preserving the structural control established by the catalog-con\-strained pipeline.

Taken together, these results show that SFT learns the structured recommendation format but does not provide reliable list-level control. Prefix-tree constrained decoding eliminates out-of-catalog generation, while the catalog-con\-strained training configuration provides consistent list-level outputs. Prompt augmentation preserves this structural reliability while improving recommendation utility, reducing repeated-item behavior, and decreasing sensitivity to prompt formulation. We therefore use the catalog-con\-strained, prompt-augmented pipeline as the controlled backbone for the remaining preprocessing and demographic-sensitivity experiments.

\begin{table*}[t]
\centering
\caption{Pipeline reliability, recommendation quality, and prompt consistency. Arrows give the preferred direction.}
\label{tab:rq1-reliability}
\small
\begin{tabular}{lrrrrrr}
\toprule
Model & HR@10 $\uparrow$ & Exact-10 $\uparrow$ & OOD@10 $\downarrow$ & Repeat@10 $\downarrow$ & Prompt HR Spread $\downarrow$ & Prompt Jaccard $\uparrow$ \\
\midrule
SFT (unconstrained decoding) & 7.47 & 0.06 & 2.08 & 3.12 & -- & -- \\
SFT (constrained decoding) & 7.43 & 0.02 & 0.00 & 4.86 & -- & -- \\
Catalog-constrained & 9.89 & 100.00 & 0.00 & 10.56 & 2.18 & 0.6799 \\
Prompt-augmented & \textbf{10.67} & 100.00 & 0.00 & \textbf{8.01} & \textbf{0.26} & \textbf{0.7489} \\
\bottomrule
\end{tabular}
\end{table*}

\subsection{RQ2: How does item-title representation affect recommendation utility and catalog usage within the catalog-con\-strained recommendation pipeline?}

RQ2 examines whether the textual representation of catalog items affects recommendation utility and catalog usage while the underlying item identities remain unchanged. The prompt-augmented configuration uses normalized titles, where articles stored at the end of MovieLens titles are moved to the beginning; for example, \textit{American President, The (1995)} becomes \textit{The American President (1995)}. The original-title condition preserves the MovieLens title format. Both conditions use the same underlying catalog and recommendation setting.

Table~\ref{tab:rq2-title-representation} shows higher recommendation utility and broader catalog usage with the original MovieLens titles. HR@10 increases from 10.67\% to 12.47\%, while Coverage@10 rises from 41.36\% to 47.88\%. The number of unique recommended items also increases from 1,606 to 1,859. Repeat@10 changes only slightly, from 8.01\% to 8.43\%.

One possible explanation for these differences is the token-prefix structure induced by the two title representations. Moving common articles to the beginning can cause more normalized titles to share high-frequency initial tokens, increasing prefix sharing at early levels of the catalog trie; \textit{The} is a common example. In contrast, the original MovieLens representation distributes titles across a wider set of initial prefixes. This changes the distribution of early prefixes and, consequently, the branching structure of the catalog prefix tree used during constrained decoding. Prior work also reported that textual item representation, and the resulting trie structure, can affect constrained generation and recommendation performance \cite{liu2026beyond,liao2026eliminating}. However, changing the title representation also changes the token sequences used during fine-tuning. Therefore, the observed differences cannot be attributed to the prefix-tree structure alone.

\begin{table}[!t]
\centering
\caption{Effect of item-title representation on recommendation utility and catalog usage. Arrows give the preferred direction.}
\label{tab:rq2-title-representation}
\small
\setlength{\tabcolsep}{2.5pt}

\begin{tabular*}{\columnwidth}{
@{\extracolsep{\fill}}
>{\raggedright\arraybackslash}p{0.27\columnwidth}
rrrr
@{}
}
\toprule
Model
& HR@10 $\uparrow$
& \shortstack{Coverage\\@10 $\uparrow$}
& \shortstack{Repeat\\@10 $\downarrow$}
& \shortstack{Unique\\Items $\uparrow$} \\
\midrule
Normalized titles
& 10.67
& 41.36
& \textbf{8.01}
& 1,606 \\

Original titles
& \textbf{12.47}
& \textbf{47.88}
& 8.43
& \textbf{1,859} \\
\bottomrule
\end{tabular*}
\end{table}

Overall, RQ2 shows that alternative title representations of the same catalog are not interchangeable. Even when the underlying item identities remain unchanged, token ordering affects recommendation utility, catalog coverage, and the range of items recommended. Therefore, title formatting should be treated as a preprocessing choice to report explicitly rather than an implementation detail.

\subsection{RQ3: How does demographic balancing in fine-tuning data affect recommendation utility and sensitivity to demographic descriptors?}

RQ3 examines how balancing the fine-tuning data with respect to gender or age affects recommendation utility and fairness-related behavior. Both balancing interventions are evaluated under the same original-title representation.
We report HR@10 to measure recommendation utility and Neutral Gap to measure the difference in HR@10 between demographic groups under the neutral prompt. Demographic sensitivity is evaluated from three complementary perspectives. Neutral-conditioned Jaccard measures the overlap between recommendations generated with the neutral prompt and the corresponding demographic-conditioned prompt. SNSR measures how much this neutral-to-conditioned similarity differs between the demographic groups. Pair Jaccard directly compares recommendations generated for the same interaction history under the two values of the evaluated demographic descriptor, such as male versus female or age $<35$ versus age $\geq35$. Higher Jaccard values indicate greater recommendation stability under demographic prompting, while lower Neutral Gap and SNSR indicate smaller group-level utility differences and smaller differences in demographic sensitivity, respectively. Table~\ref{tab:rq3-demographic-balancing} reports the results for both balancing interventions.

For gender, balancing decreases HR@10 from 12.47\% to 11.32\%. The Neutral Gap increases from 0.45 to 1.21 percentage points, indicating a larger difference in recommendation utility between the gender groups under the neutral prompt. Neutral-conditioned Jaccard decreases from 0.8312 to 0.8240, while SNSR increases from 0.0519 to 0.0636. Pair Jaccard also decreases, from 0.8110 to 0.7918. Across the reported metrics, gender-balanced fine-tuning therefore does not reduce the observed gender-related differences or sensitivity and is accompanied by lower overall recommendation utility.

The age-balanced configuration shows a different pattern. HR@10 changes only slightly, from 12.47\% to 12.39\%. The Neutral Gap decreases from 0.69 to 0.38 percentage points, while Neutral-conditioned Jaccard increases from 0.8195 to 0.8305. Pair Jaccard also increases, from 0.8407 to 0.8471. These changes indicate a smaller neutral-prompt utility difference between the age groups and greater recommendation overlap under both neutral-to-conditioned and direct descriptor-pair comparisons. However, SNSR increases from 0.0017 to 0.0077, showing that the neutral-to-conditioned similarity becomes somewhat less uniform across the two age groups. Age balancing therefore improves several of the reported age-related indicators while not improving all sensitivity measures.
\begin{table}[!t]
\centering
\caption{Effect of gender- and age-balanced fine-tuning on recommendation utility and demographic sensitivity under the original-title representation. Arrows give the preferred direction.}
\label{tab:rq3-demographic-balancing}
\small
\setlength{\tabcolsep}{2pt}

\begin{tabular*}{\columnwidth}{
@{\extracolsep{\fill}}
>{\raggedright\arraybackslash}p{0.27\columnwidth}
rrrrr
@{}
}
\toprule
Model
& HR@10 $\uparrow$
& \shortstack{Neutral\\Gap $\downarrow$}
& \shortstack{Neutral-cond.\\Jaccard $\uparrow$}
& SNSR $\downarrow$
& \shortstack{Pair\\Jaccard $\uparrow$} \\
\midrule

\multicolumn{6}{l}{\textit{Gender}} \\

Baseline
& \textbf{12.47}
& \textbf{0.45}
& \textbf{0.8312}
& \textbf{0.0519}
& \textbf{0.8110} \\

+ Gender-balanced
& 11.32
& 1.21
& 0.8240
& 0.0636
& 0.7918 \\

\midrule

\multicolumn{6}{l}{\textit{Age}} \\

Baseline
& \textbf{12.47}
& 0.69
& 0.8195
& \textbf{0.0017}
& 0.8407 \\

+ Age-balanced
& 12.39
& \textbf{0.38}
& \textbf{0.8305}
& 0.0077
& \textbf{0.8471} \\

\bottomrule
\end{tabular*}
\end{table}

Taken together, the results show that the effect of demographic balancing depends on the demographic attribute and the evaluation measure considered. Gender balancing reduces overall recommendation utility and does not improve any of the reported gender-related indicators. In contrast, age balancing largely preserves overall recommendation utility while reducing the neutral age-group utility gap and increasing both Neutral-conditioned and Pair Jaccard. At the same time, its higher SNSR shows that the improvement is not consistent across all sensitivity measures. These results indicate that demographic balancing should be evaluated separately for each demographic attribute rather than assumed to have a uniform effect on fairness-related recommendation behavior.

\subsection{RQ4: Can knowledge-enriched fine-tuning data reduce demographic sensitivity while preserving recommendation utility?}

RQ4 examines how incorporating additional item information through auxiliary fine-tuning examples affects recommendation utility and demographic sensitivity. We evaluate knowledge enrichment using the same original-title backbone and the same core evaluation dimensions as in RQ3: HR@10, Neutral Gap, Neutral-conditioned Jaccard, and SNSR.
The knowledge-enriched condition adds auxiliary examples to the fine-tuning corpus rather than replacing recommendation examples; consequently, this condition also contains more fine-tuning examples than the corresponding baseline.

Table~\ref{tab:rq4-knowledge} shows that knowledge enrichment does not substantially harm recommendation utility. Overall HR@10 decreases by only 0.23 percentage points, indicating that the main recommendation performance is largely preserved.

Knowledge enrichment reduces the Neutral Gap for both gender and age, while Neutral-conditioned Jaccard remains nearly unchanged. Thus, the smaller utility gaps are not accompanied by greater similarity between neutral and attribute-conditioned recommendation lists. This suggests that knowledge enrichment improves group-level utility alignment while leaving neutral-to-conditioned list overlap largely stable.
However, the effect on SNSR differs across attributes. Gender SNSR decreases from 0.0519 to 0.0291, whereas age SNSR increases from 0.0017 to 0.0123. Although SNSR for the age attribute remains small, the opposite directions show that knowledge enrichment does not affect between-group differences in demographic sensitivity uniformly across attributes.

\begin{table}[!t]
\centering
\caption{Effect of knowledge enrichment on recommendation utility and demographic sensitivity. Arrows give the preferred direction.}
\label{tab:rq4-knowledge}
\small
\setlength{\tabcolsep}{2.5pt}

\begin{tabular*}{\columnwidth}{
@{\extracolsep{\fill}}
lrrrr
@{}
}
\toprule
Model
& HR@10 $\uparrow$
& \shortstack{Neutral\\Gap $\downarrow$}
& \shortstack{Neutral-cond.\\Jaccard $\uparrow$}
& SNSR $\downarrow$ \\
\midrule

\multicolumn{5}{l}{\textit{Gender}} \\
Baseline
& \textbf{12.47}
& 0.45
& \textbf{0.8312}
& 0.0519 \\
+ Knowledge
& 12.24
& \textbf{0.10}
& 0.8270
& \textbf{0.0291} \\

\midrule

\multicolumn{5}{l}{\textit{Age}} \\
Baseline
& \textbf{12.47}
& 0.69
& 0.8195
& \textbf{0.0017} \\
+ Knowledge
& 12.24
& \textbf{0.32}
& \textbf{0.8201}
& 0.0123 \\

\bottomrule
\end{tabular*}
\end{table}

Overall, knowledge enrichment in fine-tuning data largely preserves recommendation utility while reducing group-level utility gaps for both gender and age. Its effect on SNSR is attribute-dependent: it lowers gender-related sensitivity but raises the age-related value, which nevertheless remains small in absolute terms. As in RQ3, the fairness-related effect of this preprocessing choice should therefore be interpreted separately for each attribute rather than assumed to be uniform. These findings suggest that incorporating additional item information into fine-tuning data can serve as a complementary data-level mechanism for reducing some forms of demographic sensitivity, even without an explicit fairness-specific training objective.

\section{Conclusion}

In this study, we examined how preprocessing choices in fine-tuning data affect utility and fairness-related behavior in generative LLM-based recommendation. Catalog-constrained decoding removed out-of-catalog items, the catalog-con\-strained training configuration produced complete ten-item outputs, and prompt augmentation improved prompt consistency and utility. Item-title representation affected both utility and catalog coverage even though the underlying item identities remained unchanged.
The effects of demographic balancing and auxiliary item-knowledge enrichment were attribute-dependent. Age balancing reduced the age-related utility gap while largely preserving utility, whereas gender balancing reduced utility without improving the evaluated gender-related indicators. Knowledge enrichment preserved utility and reduced group-level gaps, but its effect on demographic sensitivity varied across attributes and measures. A preprocessing choice therefore cannot be judged on utility alone, nor on a single attribute. Future work should control the size of the knowledge-enriched corpus and extend the evaluation beyond one dataset and binary demographic groups.

\begin{acks}
This work is partially supported by the Scientific Research Project Unit of Istanbul Technical University under project number: 46051.
\end{acks}

\bibliographystyle{ACM-Reference-Format}
\bibliography{sample-base}


\begin{thebibliography}{37}


\ifx \showCODEN    \undefined \def \showCODEN     #1{\unskip}     \fi
\ifx \showISBNx    \undefined \def \showISBNx     #1{\unskip}     \fi
\ifx \showISBNxiii \undefined \def \showISBNxiii  #1{\unskip}     \fi
\ifx \showISSN     \undefined \def \showISSN      #1{\unskip}     \fi
\ifx \showLCCN     \undefined \def \showLCCN      #1{\unskip}     \fi
\ifx \shownote     \undefined \def \shownote      #1{#1}          \fi
\ifx \showarticletitle \undefined \def \showarticletitle #1{#1}   \fi
\ifx \showURL      \undefined \def \showURL       {\relax}        \fi
\providecommand\bibfield[2]{#2}
\providecommand\bibinfo[2]{#2}
\providecommand\natexlab[1]{#1}
\providecommand\showeprint[2][]{arXiv:#2}

\bibitem[Andre et~al\mbox{.}(2025)]%
        {andre2025bias}
\bibfield{author}{\bibinfo{person}{Alexandre Andre}, \bibinfo{person}{Gauthier
  Roy}, \bibinfo{person}{Eva Dyer}, {and} \bibinfo{person}{Kai Wang}.}
  \bibinfo{year}{2025}\natexlab{}.
\newblock \showarticletitle{Revealing Potential Biases in {LLM}-Based
  Recommender Systems in the Cold Start Setting}.
\newblock \bibinfo{journal}{\emph{arXiv preprint arXiv:2508.20401}}
  (\bibinfo{year}{2025}).
\newblock


\bibitem[Bao et~al\mbox{.}(2023)]%
        {bao2023tallrec}
\bibfield{author}{\bibinfo{person}{Keqin Bao}, \bibinfo{person}{Jizhi Zhang},
  \bibinfo{person}{Yang Zhang}, \bibinfo{person}{Wenjie Wang},
  \bibinfo{person}{Fuli Feng}, {and} \bibinfo{person}{Xiangnan He}.}
  \bibinfo{year}{2023}\natexlab{}.
\newblock \showarticletitle{{TALLRec}: An effective and efficient tuning
  framework to align large language model with recommendation}. In
  \bibinfo{booktitle}{\emph{Proceedings of the 17th ACM conference on
  recommender systems}}. \bibinfo{pages}{1007--1014}.
\newblock


\bibitem[Cao and Lio(2024)]%
        {cao2024genrec}
\bibfield{author}{\bibinfo{person}{Panfeng Cao} {and} \bibinfo{person}{Pietro
  Lio}.} \bibinfo{year}{2024}\natexlab{}.
\newblock \showarticletitle{{GenRec}: generative sequential recommendation with
  large language models}.
\newblock \bibinfo{journal}{\emph{arXiv preprint arXiv:2407.21191}}
  (\bibinfo{year}{2024}).
\newblock


\bibitem[Chatterjee et~al\mbox{.}(2024)]%
        {chatterjee-etal-2024-posix}
\bibfield{author}{\bibinfo{person}{Anwoy Chatterjee}, \bibinfo{person}{H~S V N
  S~Kowndinya Renduchintala}, \bibinfo{person}{Sumit Bhatia}, {and}
  \bibinfo{person}{Tanmoy Chakraborty}.} \bibinfo{year}{2024}\natexlab{}.
\newblock \showarticletitle{{POSIX}: A Prompt Sensitivity Index For Large
  Language Models}. In \bibinfo{booktitle}{\emph{Findings of the Association
  for Computational Linguistics: EMNLP 2024}},
  \bibfield{editor}{\bibinfo{person}{Yaser Al-Onaizan}, \bibinfo{person}{Mohit
  Bansal}, {and} \bibinfo{person}{Yun-Nung Chen}} (Eds.).
  \bibinfo{publisher}{Association for Computational Linguistics},
  \bibinfo{address}{Miami, Florida, USA}, \bibinfo{pages}{14550--14565}.
\newblock
\href{https://doi.org/10.18653/v1/2024.findings-emnlp.852}{doi:\nolinkurl{10.18653/v1/2024.findings-emnlp.852}}


\bibitem[Chen et~al\mbox{.}(2023)]%
        {chen2023improving}
\bibfield{author}{\bibinfo{person}{Lei Chen}, \bibinfo{person}{Le Wu},
  \bibinfo{person}{Kun Zhang}, \bibinfo{person}{Richang Hong},
  \bibinfo{person}{Defu Lian}, \bibinfo{person}{Zhiqiang Zhang},
  \bibinfo{person}{Jun Zhou}, {and} \bibinfo{person}{Meng Wang}.}
  \bibinfo{year}{2023}\natexlab{}.
\newblock \showarticletitle{Improving recommendation fairness via data
  augmentation}. In \bibinfo{booktitle}{\emph{Proceedings of the ACM web
  conference 2023}}. \bibinfo{pages}{1012--1020}.
\newblock


\bibitem[Dettmers et~al\mbox{.}(2023)]%
        {dettmers2023qlora}
\bibfield{author}{\bibinfo{person}{Tim Dettmers}, \bibinfo{person}{Artidoro
  Pagnoni}, \bibinfo{person}{Ari Holtzman}, {and} \bibinfo{person}{Luke
  Zettlemoyer}.} \bibinfo{year}{2023}\natexlab{}.
\newblock \showarticletitle{{QLoRA}: Efficient Finetuning of Quantized {LLMs}}.
  In \bibinfo{booktitle}{\emph{Advances in Neural Information Processing
  Systems}}, Vol.~\bibinfo{volume}{36}.
\newblock


\bibitem[Fayyazi et~al\mbox{.}(2025)]%
        {fayyazi2025facter}
\bibfield{author}{\bibinfo{person}{Arya Fayyazi}, \bibinfo{person}{Mehdi
  Kamal}, {and} \bibinfo{person}{Massoud Pedram}.}
  \bibinfo{year}{2025}\natexlab{}.
\newblock \showarticletitle{{FACTER}: Fairness-aware conformal thresholding and
  prompt engineering for enabling fair {LLM}-based recommender systems}.
\newblock \bibinfo{journal}{\emph{arXiv preprint arXiv:2502.02966}}
  (\bibinfo{year}{2025}).
\newblock


\bibitem[Geng et~al\mbox{.}(2022)]%
        {geng2022p5}
\bibfield{author}{\bibinfo{person}{Shijie Geng}, \bibinfo{person}{Shuchang
  Liu}, \bibinfo{person}{Zuohui Fu}, \bibinfo{person}{Yingqiang Ge}, {and}
  \bibinfo{person}{Yongfeng Zhang}.} \bibinfo{year}{2022}\natexlab{}.
\newblock \showarticletitle{Recommendation as Language Processing ({RLP}): A
  Unified Pretrain, Personalized Prompt \& Predict Paradigm ({P5})}.
\newblock \bibinfo{journal}{\emph{arXiv preprint arXiv:2203.13366}}
  (\bibinfo{year}{2022}).
\newblock


\bibitem[Grattafiori et~al\mbox{.}(2024)]%
        {llama2024herd}
\bibfield{author}{\bibinfo{person}{Aaron Grattafiori} {et~al\mbox{.}}}
  \bibinfo{year}{2024}\natexlab{}.
\newblock \showarticletitle{The {Llama} 3 Herd of Models}.
\newblock \bibinfo{journal}{\emph{arXiv preprint arXiv:2407.21783}}
  (\bibinfo{year}{2024}).
\newblock


\bibitem[Harper and Konstan(2015)]%
        {harper2015movielens}
\bibfield{author}{\bibinfo{person}{F~Maxwell Harper} {and}
  \bibinfo{person}{Joseph~A Konstan}.} \bibinfo{year}{2015}\natexlab{}.
\newblock \showarticletitle{The {MovieLens} datasets: History and context}.
\newblock \bibinfo{journal}{\emph{{ACM} Transactions on Interactive Intelligent
  Systems ({TiiS})}} \bibinfo{volume}{5}, \bibinfo{number}{4}
  (\bibinfo{year}{2015}), \bibinfo{pages}{1--19}.
\newblock


\bibitem[Hou et~al\mbox{.}(2026)]%
        {hou2026reccocktail}
\bibfield{author}{\bibinfo{person}{Min Hou}, \bibinfo{person}{Chenxi Bai},
  \bibinfo{person}{Le Wu}, \bibinfo{person}{Hao Liu}, \bibinfo{person}{Kai
  Zhang}, \bibinfo{person}{Weiwen Liu}, \bibinfo{person}{Richang Hong},
  \bibinfo{person}{Ruiming Tang}, {and} \bibinfo{person}{Meng Wang}.}
  \bibinfo{year}{2026}\natexlab{}.
\newblock \showarticletitle{{RecCocktail}: A Generalizable and Efficient
  Framework for {LLM}-Based Recommendation}. In
  \bibinfo{booktitle}{\emph{Proceedings of the AAAI Conference on Artificial
  Intelligence}}, Vol.~\bibinfo{volume}{40}. \bibinfo{pages}{14838--14847}.
\newblock


\bibitem[Hu et~al\mbox{.}(2022)]%
        {hu2022lora}
\bibfield{author}{\bibinfo{person}{Edward~J Hu}, \bibinfo{person}{Yelong Shen},
  \bibinfo{person}{Phillip Wallis}, \bibinfo{person}{Zeyuan Allen-Zhu},
  \bibinfo{person}{Yuanzhi Li}, \bibinfo{person}{Shean Wang},
  \bibinfo{person}{Lu Wang}, {and} \bibinfo{person}{Weizhu Chen}.}
  \bibinfo{year}{2022}\natexlab{}.
\newblock \showarticletitle{{LoRA}: Low-Rank Adaptation of Large Language
  Models}. In \bibinfo{booktitle}{\emph{The Tenth International Conference on
  Learning Representations}}. \bibinfo{publisher}{OpenReview.net}.
\newblock
\urldef\tempurl%
\url{https://openreview.net/forum?id=nZeVKeeFYf9}
\showURL{%
\tempurl}


\bibitem[Hua et~al\mbox{.}(2024)]%
        {hua2024up5}
\bibfield{author}{\bibinfo{person}{Wenyue Hua}, \bibinfo{person}{Yingqiang Ge},
  \bibinfo{person}{Shuyuan Xu}, \bibinfo{person}{Jianchao Ji},
  \bibinfo{person}{Zelong Li}, {and} \bibinfo{person}{Yongfeng Zhang}.}
  \bibinfo{year}{2024}\natexlab{}.
\newblock \showarticletitle{{UP5}: Unbiased foundation model for fairness-aware
  recommendation}. In \bibinfo{booktitle}{\emph{Proceedings of the 18th
  Conference of the European Chapter of the Association for Computational
  Linguistics (Volume 1: Long Papers)}}. \bibinfo{pages}{1899--1912}.
\newblock


\bibitem[Kweon et~al\mbox{.}(2025)]%
        {kweon2025uncertainty}
\bibfield{author}{\bibinfo{person}{Wonbin Kweon}, \bibinfo{person}{Sanghwan
  Jang}, \bibinfo{person}{SeongKu Kang}, {and} \bibinfo{person}{Hwanjo Yu}.}
  \bibinfo{year}{2025}\natexlab{}.
\newblock \showarticletitle{Uncertainty Quantification and Decomposition for
  {LLM}-Based Recommendation}. In \bibinfo{booktitle}{\emph{Proceedings of the
  ACM Web Conference 2025}}. \bibinfo{pages}{4889--4901}.
\newblock
\href{https://doi.org/10.1145/3696410.3714601}{doi:\nolinkurl{10.1145/3696410.3714601}}


\bibitem[Kwon et~al\mbox{.}(2023)]%
        {kwon2023vllm}
\bibfield{author}{\bibinfo{person}{Woosuk Kwon}, \bibinfo{person}{Zhuohan Li},
  \bibinfo{person}{Siyuan Zhuang}, \bibinfo{person}{Ying Sheng},
  \bibinfo{person}{Lianmin Zheng}, \bibinfo{person}{Cody~Hao Yu},
  \bibinfo{person}{Joseph~E. Gonzalez}, \bibinfo{person}{Hao Zhang}, {and}
  \bibinfo{person}{Ion Stoica}.} \bibinfo{year}{2023}\natexlab{}.
\newblock \showarticletitle{Efficient Memory Management for Large Language
  Model Serving with {PagedAttention}}. In
  \bibinfo{booktitle}{\emph{Proceedings of the 29th Symposium on Operating
  Systems Principles}}. \bibinfo{pages}{611--626}.
\newblock
\href{https://doi.org/10.1145/3600006.3613165}{doi:\nolinkurl{10.1145/3600006.3613165}}


\bibitem[Li et~al\mbox{.}(2025a)]%
        {li-etal-2025-chatcrs}
\bibfield{author}{\bibinfo{person}{Chuang Li}, \bibinfo{person}{Yang Deng},
  \bibinfo{person}{Hengchang Hu}, \bibinfo{person}{Min-Yen Kan}, {and}
  \bibinfo{person}{Haizhou Li}.} \bibinfo{year}{2025}\natexlab{a}.
\newblock \showarticletitle{{ChatCRS}: Incorporating external knowledge and
  goal guidance for {LLM}-based conversational recommender systems}. In
  \bibinfo{booktitle}{\emph{Findings of the association for computational
  linguistics: NAACL 2025}}. \bibinfo{pages}{295--312}.
\newblock


\bibitem[Li et~al\mbox{.}(2025b)]%
        {li2025ganprompt}
\bibfield{author}{\bibinfo{person}{Xinyu Li}, \bibinfo{person}{Chuang Zhao},
  \bibinfo{person}{Hongke Zhao}, \bibinfo{person}{Likang Wu},
  \bibinfo{person}{Ming He}, {and} \bibinfo{person}{Jianping Fan}.}
  \bibinfo{year}{2025}\natexlab{b}.
\newblock \showarticletitle{{GANPrompt}: Improving {LLM}-Based Recommendations
  with {GAN}-Enhanced Diversity Prompts}.
\newblock \bibinfo{journal}{\emph{ACM Transactions on Information Systems}}
  \bibinfo{volume}{44}, \bibinfo{number}{1} (\bibinfo{year}{2025}),
  \bibinfo{pages}{1--36}.
\newblock


\bibitem[Liang et~al\mbox{.}(2025)]%
        {liang-etal-2025-taxonomy}
\bibfield{author}{\bibinfo{person}{Yueqing Liang}, \bibinfo{person}{Liangwei
  Yang}, \bibinfo{person}{Chen Wang}, \bibinfo{person}{Xiongxiao Xu},
  \bibinfo{person}{Philip~S Yu}, {and} \bibinfo{person}{Kai Shu}.}
  \bibinfo{year}{2025}\natexlab{}.
\newblock \showarticletitle{Taxonomy-guided zero-shot recommendations with
  {LLMs}}. In \bibinfo{booktitle}{\emph{Proceedings of the 31st International
  Conference on Computational Linguistics}}. \bibinfo{pages}{1520--1530}.
\newblock


\bibitem[Liao et~al\mbox{.}(2026)]%
        {liao2026eliminating}
\bibfield{author}{\bibinfo{person}{Hao Liao}, \bibinfo{person}{Jiwei Zhang},
  \bibinfo{person}{Jianxun Lian}, \bibinfo{person}{Wensheng Lu},
  \bibinfo{person}{Mingqi Wu}, \bibinfo{person}{Yong Zhang},
  \bibinfo{person}{Yitian Huang}, \bibinfo{person}{Mingyang Zhou},
  \bibinfo{person}{Rui Mao}, {et~al\mbox{.}}} \bibinfo{year}{2026}\natexlab{}.
\newblock \showarticletitle{Eliminating Out-of-Domain Recommendations in
  {LLM}-based Recommender Systems: A Unified View}. In
  \bibinfo{booktitle}{\emph{Findings of the Association for Computational
  Linguistics: ACL 2026}}. \bibinfo{pages}{6251--6271}.
\newblock


\bibitem[Liao et~al\mbox{.}(2024)]%
        {liao2024llara}
\bibfield{author}{\bibinfo{person}{Jiayi Liao}, \bibinfo{person}{Sihang Li},
  \bibinfo{person}{Zhengyi Yang}, \bibinfo{person}{Jiancan Wu},
  \bibinfo{person}{Yancheng Yuan}, \bibinfo{person}{Xiang Wang}, {and}
  \bibinfo{person}{Xiangnan He}.} \bibinfo{year}{2024}\natexlab{}.
\newblock \showarticletitle{{LLaRA}: Large language-recommendation assistant}.
  In \bibinfo{booktitle}{\emph{Proceedings of the 47th international ACM SIGIR
  conference on research and development in information retrieval}}.
  \bibinfo{pages}{1785--1795}.
\newblock


\bibitem[Lin et~al\mbox{.}(2025)]%
        {lin2025can}
\bibfield{author}{\bibinfo{person}{Jianghao Lin}, \bibinfo{person}{Xinyi Dai},
  \bibinfo{person}{Yunjia Xi}, \bibinfo{person}{Weiwen Liu},
  \bibinfo{person}{Bo Chen}, \bibinfo{person}{Hao Zhang}, \bibinfo{person}{Yong
  Liu}, \bibinfo{person}{Chuhan Wu}, \bibinfo{person}{Xiangyang Li},
  \bibinfo{person}{Chenxu Zhu}, {et~al\mbox{.}}}
  \bibinfo{year}{2025}\natexlab{}.
\newblock \showarticletitle{How can recommender systems benefit from large
  language models: A survey}.
\newblock \bibinfo{journal}{\emph{ACM Transactions on Information Systems}}
  \bibinfo{volume}{43}, \bibinfo{number}{2} (\bibinfo{year}{2025}),
  \bibinfo{pages}{1--47}.
\newblock


\bibitem[Lin et~al\mbox{.}(2024)]%
        {lin2024data}
\bibfield{author}{\bibinfo{person}{Xinyu Lin}, \bibinfo{person}{Wenjie Wang},
  \bibinfo{person}{Yongqi Li}, \bibinfo{person}{Shuo Yang},
  \bibinfo{person}{Fuli Feng}, \bibinfo{person}{Yinwei Wei}, {and}
  \bibinfo{person}{Tat-Seng Chua}.} \bibinfo{year}{2024}\natexlab{}.
\newblock \showarticletitle{Data-efficient Fine-tuning for {LLM}-based
  Recommendation}. In \bibinfo{booktitle}{\emph{Proceedings of the 47th
  international ACM SIGIR conference on research and development in information
  retrieval}}. \bibinfo{pages}{365--374}.
\newblock


\bibitem[Liu et~al\mbox{.}(2024)]%
        {liu2024p2rec}
\bibfield{author}{\bibinfo{person}{Dugang Liu}, \bibinfo{person}{Shenxian
  Xian}, \bibinfo{person}{Xiaolin Lin}, \bibinfo{person}{Xiaolian Zhang},
  \bibinfo{person}{Hong Zhu}, \bibinfo{person}{Yuan Fang},
  \bibinfo{person}{Zhen Chen}, {and} \bibinfo{person}{Zhong Ming}.}
  \bibinfo{year}{2024}\natexlab{}.
\newblock \showarticletitle{A Practice-Friendly {LLM}-Enhanced Paradigm with
  Preference Parsing for Sequential Recommendation}.
\newblock \bibinfo{journal}{\emph{arXiv preprint arXiv:2406.00333}}
  (\bibinfo{year}{2024}).
\newblock


\bibitem[Liu et~al\mbox{.}(2026)]%
        {liu2026beyond}
\bibfield{author}{\bibinfo{person}{Jingzhe Liu}, \bibinfo{person}{Hanbing
  Wang}, \bibinfo{person}{Jiliang Tang}, \bibinfo{person}{Liam Collins},
  \bibinfo{person}{Tong Zhao}, \bibinfo{person}{Neil Shah}, {and}
  \bibinfo{person}{Mingxuan Ju}.} \bibinfo{year}{2026}\natexlab{}.
\newblock \showarticletitle{Beyond Fixed Depths and Widths: Optimizing Textual
  Decoding Tries in {LLM}-based Generative Recommendation}.
\newblock \bibinfo{journal}{\emph{arXiv preprint arXiv:2607.16633}}
  (\bibinfo{year}{2026}).
\newblock


\bibitem[Melchiorre et~al\mbox{.}(2021)]%
        {melchiorre2021investigating}
\bibfield{author}{\bibinfo{person}{Alessandro~B Melchiorre},
  \bibinfo{person}{Navid Rekabsaz}, \bibinfo{person}{Emilia Parada-Cabaleiro},
  \bibinfo{person}{Stefan Brandl}, \bibinfo{person}{Oleg Lesota}, {and}
  \bibinfo{person}{Markus Schedl}.} \bibinfo{year}{2021}\natexlab{}.
\newblock \showarticletitle{Investigating gender fairness of recommendation
  algorithms in the music domain}.
\newblock \bibinfo{journal}{\emph{Information Processing \& Management}}
  \bibinfo{volume}{58}, \bibinfo{number}{5} (\bibinfo{year}{2021}),
  \bibinfo{pages}{102666}.
\newblock


\bibitem[Mizrahi et~al\mbox{.}(2024)]%
        {mizrahi-etal-2024-state}
\bibfield{author}{\bibinfo{person}{Moran Mizrahi}, \bibinfo{person}{Guy
  Kaplan}, \bibinfo{person}{Dan Malkin}, \bibinfo{person}{Rotem Dror},
  \bibinfo{person}{Dafna Shahaf}, {and} \bibinfo{person}{Gabriel Stanovsky}.}
  \bibinfo{year}{2024}\natexlab{}.
\newblock \showarticletitle{State of What Art? A Call for Multi-Prompt {LLM}
  Evaluation}.
\newblock \bibinfo{journal}{\emph{Transactions of the Association for
  Computational Linguistics}}  \bibinfo{volume}{12} (\bibinfo{year}{2024}),
  \bibinfo{pages}{933--949}.
\newblock
\href{https://doi.org/10.1162/tacl_a_00681}{doi:\nolinkurl{10.1162/tacl_a_00681}}


\bibitem[Munson et~al\mbox{.}(2025)]%
        {munson2025review}
\bibfield{author}{\bibinfo{person}{Jacob Munson}, \bibinfo{person}{Thomas
  Cuezze}, \bibinfo{person}{Siddat Nesar}, {and} \bibinfo{person}{Dominique
  Zosso}.} \bibinfo{year}{2025}\natexlab{}.
\newblock \showarticletitle{A review of large language models and the
  recommendation task}.
\newblock \bibinfo{journal}{\emph{Discover Artificial Intelligence}}
  \bibinfo{volume}{5}, \bibinfo{number}{1} (\bibinfo{year}{2025}),
  \bibinfo{pages}{203}.
\newblock


\bibitem[Petrov and Macdonald(2023)]%
        {petrov2023gptrec}
\bibfield{author}{\bibinfo{person}{Aleksandr~V. Petrov} {and}
  \bibinfo{person}{Craig Macdonald}.} \bibinfo{year}{2023}\natexlab{}.
\newblock \showarticletitle{Generative Sequential Recommendation with
  {GPTRec}}.
\newblock \bibinfo{journal}{\emph{arXiv preprint arXiv:2306.11114}}
  (\bibinfo{year}{2023}).
\newblock


\bibitem[Qiang et~al\mbox{.}(2024)]%
        {qiang-etal-2024-prompt}
\bibfield{author}{\bibinfo{person}{Yao Qiang}, \bibinfo{person}{Subhrangshu
  Nandi}, \bibinfo{person}{Ninareh Mehrabi}, \bibinfo{person}{Greg Ver~Steeg},
  \bibinfo{person}{Anoop Kumar}, \bibinfo{person}{Anna Rumshisky}, {and}
  \bibinfo{person}{Aram Galstyan}.} \bibinfo{year}{2024}\natexlab{}.
\newblock \showarticletitle{Prompt Perturbation Consistency Learning for Robust
  Language Models}. In \bibinfo{booktitle}{\emph{Findings of the Association
  for Computational Linguistics: EACL 2024}},
  \bibfield{editor}{\bibinfo{person}{Yvette Graham} {and}
  \bibinfo{person}{Matthew Purver}} (Eds.). \bibinfo{publisher}{Association for
  Computational Linguistics}, \bibinfo{address}{St. Julian{'}s, Malta},
  \bibinfo{pages}{1357--1370}.
\newblock
\href{https://doi.org/10.18653/v1/2024.findings-eacl.91}{doi:\nolinkurl{10.18653/v1/2024.findings-eacl.91}}


\bibitem[Sah et~al\mbox{.}(2025)]%
        {sah2025faireval}
\bibfield{author}{\bibinfo{person}{Chandan~Kumar Sah}, \bibinfo{person}{Xiaoli
  Lian}, \bibinfo{person}{Tony Xu}, {and} \bibinfo{person}{Li Zhang}.}
  \bibinfo{year}{2025}\natexlab{}.
\newblock \showarticletitle{{FairEval}: Evaluating fairness in {LLM}-based
  recommendations with personality awareness}.
\newblock \bibinfo{journal}{\emph{arXiv preprint arXiv:2504.07801}}
  (\bibinfo{year}{2025}).
\newblock


\bibitem[Vass{\o}y and Langseth(2024)]%
        {vassoy2024consumer}
\bibfield{author}{\bibinfo{person}{Bj{\o}rnar Vass{\o}y} {and}
  \bibinfo{person}{Helge Langseth}.} \bibinfo{year}{2024}\natexlab{}.
\newblock \showarticletitle{Consumer-side fairness in recommender systems: a
  systematic survey of methods and evaluation}.
\newblock \bibinfo{journal}{\emph{Artificial Intelligence Review}}
  \bibinfo{volume}{57}, \bibinfo{number}{4} (\bibinfo{year}{2024}),
  \bibinfo{pages}{101}.
\newblock


\bibitem[Wang et~al\mbox{.}(2024)]%
        {wang2024large}
\bibfield{author}{\bibinfo{person}{Jianling Wang}, \bibinfo{person}{Haokai Lu},
  \bibinfo{person}{James Caverlee}, \bibinfo{person}{Ed~H Chi}, {and}
  \bibinfo{person}{Minmin Chen}.} \bibinfo{year}{2024}\natexlab{}.
\newblock \showarticletitle{Large language models as data augmenters for
  cold-start item recommendation}. In \bibinfo{booktitle}{\emph{Companion
  proceedings of the ACM web conference 2024}}. \bibinfo{pages}{726--729}.
\newblock


\bibitem[Wang et~al\mbox{.}(2025)]%
        {wang2025beyond}
\bibfield{author}{\bibinfo{person}{Zehui Wang}, \bibinfo{person}{Wolfram
  H{\"o}pken}, {and} \bibinfo{person}{Dietmar Jannach}.}
  \bibinfo{year}{2025}\natexlab{}.
\newblock \showarticletitle{Beyond visit trajectories: enhancing {POI}
  recommendation via {LLM}-augmented text and image representations}. In
  \bibinfo{booktitle}{\emph{Proceedings of the Nineteenth ACM Conference on
  Recommender Systems}}. \bibinfo{pages}{521--526}.
\newblock


\bibitem[Wu et~al\mbox{.}(2023)]%
        {wu2023fairness}
\bibfield{author}{\bibinfo{person}{Yao Wu}, \bibinfo{person}{Jian Cao}, {and}
  \bibinfo{person}{Guandong Xu}.} \bibinfo{year}{2023}\natexlab{}.
\newblock \showarticletitle{Fairness in recommender systems: evaluation
  approaches and assurance strategies}.
\newblock \bibinfo{journal}{\emph{ACM Transactions on Knowledge Discovery from
  Data}} \bibinfo{volume}{18}, \bibinfo{number}{1} (\bibinfo{year}{2023}),
  \bibinfo{pages}{1--37}.
\newblock


\bibitem[Xu et~al\mbox{.}(2025)]%
        {xu2025rallrec}
\bibfield{author}{\bibinfo{person}{Jian Xu}, \bibinfo{person}{Sichun Luo},
  \bibinfo{person}{Xiangyu Chen}, \bibinfo{person}{Haoming Huang},
  \bibinfo{person}{Hanxu Hou}, {and} \bibinfo{person}{Linqi Song}.}
  \bibinfo{year}{2025}\natexlab{}.
\newblock \showarticletitle{{RALLRec}: Improving retrieval augmented large
  language model recommendation with representation learning}. In
  \bibinfo{booktitle}{\emph{Companion Proceedings of the ACM on Web Conference
  2025}}. \bibinfo{pages}{1436--1440}.
\newblock


\bibitem[Xu et~al\mbox{.}(2024)]%
        {xu2024openp5}
\bibfield{author}{\bibinfo{person}{Shuyuan Xu}, \bibinfo{person}{Wenyue Hua},
  {and} \bibinfo{person}{Yongfeng Zhang}.} \bibinfo{year}{2024}\natexlab{}.
\newblock \showarticletitle{{OpenP5}: An Open-Source Platform for Developing,
  Training, and Evaluating {LLM}-based Recommender Systems}.
\newblock \bibinfo{journal}{\emph{arXiv preprint arXiv:2306.11134}}
  (\bibinfo{year}{2024}).
\newblock


\bibitem[Zhang et~al\mbox{.}(2023)]%
        {zhang2023chatgptfair}
\bibfield{author}{\bibinfo{person}{Jizhi Zhang}, \bibinfo{person}{Keqin Bao},
  \bibinfo{person}{Yang Zhang}, \bibinfo{person}{Wenjie Wang},
  \bibinfo{person}{Fuli Feng}, {and} \bibinfo{person}{Xiangnan He}.}
  \bibinfo{year}{2023}\natexlab{}.
\newblock \showarticletitle{Is {ChatGPT} fair for recommendation? evaluating
  fairness in large language model recommendation}. In
  \bibinfo{booktitle}{\emph{Proceedings of the 17th ACM conference on
  recommender systems}}. \bibinfo{pages}{993--999}.
\newblock


\end{thebibliography}

\end{document}